\documentclass[aip,amsmath,amssymb,reprint]{revtex4-1}
\usepackage{graphicx}
\usepackage{dcolumn}
\usepackage{bm}
\usepackage[utf8]{inputenc}
\usepackage[T1]{fontenc}
\usepackage{mathptmx}
\usepackage{etoolbox}
\usepackage{multirow} 

\makeatletter
\def\@email#1#2{%
 \endgroup
 \patchcmd{\titleblock@produce}
  {\frontmatter@RRAPformat}
  {\frontmatter@RRAPformat{\produce@RRAP{*#1\href{mailto:#2}{#2}}}\frontmatter@RRAPformat}
  {}{}
}%
\makeatother

\usepackage{xcolor}
\usepackage[normalem]{ulem}

\newcommand{\etal}{\textit{et al.}}

\newcommand{\Ap}{\ensuremath{\text{A}^{+}}}
\newcommand{\He}[1]{\ensuremath{\text{He}_{#1}}}
\newcommand{\ApHe}[1]{\ensuremath{\Ap@\He #1}}
\newcommand{\nApHe}[2]{\ensuremath{#1\Ap@\He #2}}
\newcommand{\Nap}{\ensuremath{\text{Na}^{+}}}
\newcommand{\Calp}{\ensuremath{\text{Ca}^{+}}}

\begin{document}

\title {Revisiting Thomson's model with multiply charged superfluid helium nanodroplets}

\author{Ernesto Garc\'{\i}a-Alfonso}
\affiliation{Laboratoire Collisions, Agr\'egats, R\'eactivit\'e
(LCAR), Universit\'e de Toulouse, CNRS, 31062 Toulouse, France}
 
\author{Francesco Ancilotto}
\affiliation{Dipartimento di Fisica e Astronomia ``Galileo Galilei''
and CNISM, Universit\`a di Padova, via Marzolo 8, 35122 Padova, Italy}
\affiliation{ CNR-Istituto Officina dei Materiali (IOM), via Bonomea, 265 - 34136 Trieste, Italy }

\author{Manuel Barranco}
\affiliation{Departament FQA, Facultat de F\'{\i}sica,
Universitat de Barcelona, Av.\ Diagonal 645,
08028 Barcelona, Spain.}
\affiliation{Institute of Nanoscience and Nanotechnology (IN2UB),
Universitat de Barcelona, Barcelona, Spain.}

\author{Fausto Cargnoni}
\affiliation{CNR-SCITEC, Via Golgi 19, 20133 Milano, Italy.}

\author{Nadine Halberstadt}
\affiliation{Laboratoire Collisions, Agr\'egats, R\'eactivit\'e
(LCAR), Universit\'e de Toulouse, CNRS, 31062, Toulouse, France}

\author{Mart\'{\i} Pi}
\affiliation{Departament FQA, Facultat de F\'{\i}sica,
Universitat de Barcelona, Av.\ Diagonal 645,
08028 Barcelona, Spain.}
\affiliation{Institute of Nanoscience and Nanotechnology (IN2UB),
Universitat de Barcelona, Barcelona, Spain.}

\begin{abstract}
We study superfluid helium droplets multiply charged withNa$^+$ or Ca$^+$  ions. 
When stable, the charges are found to reside in equilibrium close to the droplet surface, thus representing a physical realization of Thomson’s model. 
We find the minimum radius of the helium droplet that can host a given number of ions using a model whose physical ingredients are the solvation energy of the cations, calculated within the He-DFT approach, and their mutual Coulomb repulsion energy. 
Our model goes beyond the often used liquid drop model, where charges are smeared out either within the droplet or on its surface, and which neglects the solid-like helium shell around the ions. 
We find that below a threshold droplet radius $R_0$, the total energy of the system becomes higher than that of the separated system of the pristine helium droplet and the charges embedded in their solvation microcluster (“snowball”). 
However, the ions are still kept within the droplet by the presence of energy barriers which hinder Coulomb explosion. 
A further reduction of the droplet radius below a value $R_{expl}$ eventually results in the disappearance of such barrier, leading to Coulomb explosion. 
Surprisingly, our results are rather insensitive to the ion atomic species. 
This makes room to discuss them in the context of intrinsic multicharged helium droplets, where the charges are triatomic He$_3^+$ ions. 
Our calculated values for $R_{expl}$ display the correct scaling with the number of cations compared to available experimental results, at variance with other estimates for the critical radii.

\end{abstract}

\date{\today}

\maketitle

\section{Introduction}
Multiply charged helium nanodroplets have been expected to undergo a fission-like process  
due to Coulomb repulsion between charges. 
Consequently, the possibility of creating multiply charged helium droplets that
could be stable or metastable  in the millisecond time range spanned in 
usual experiments\cite{Toe04} had seldom been 
considered, except for doubly charged helium drops produced by electron impact.\cite{Far97} 
In that work, doubly positively charged  
He$^{2+}_N$ drops were found for $N$ larger than $N \sim 2 \times 10^5$.
The situation has recently changed. Using a novel experimental setup,
Laimer \etal\cite{Lai19} and Feinberg \etal\cite{Fei22} have shown  
that superfluid helium drops hosting several tens of positive charges 
can readily be formed by electron impact, and that these drops are stable 
on the millisecond timescale of the experiment. The radii $R$ of these multiply charged 
droplets were determined as a function of the charge state $n$, showing that $R\propto n^{1/2}$. 
Since these studies involved pure drops, the ions generated were intrinsic ions.
In a theoretical study using a density functional approach specific to superfluid helium (He-DFT),\cite{Mat14}
the linear, covalently-bound He$_3^+$ cluster ion was identified as the core of a local arrangement characterized 
by an important enhancement of helium density, usually called ``snowball''.
He$_3^+$ ions produced by electron impact 
in large droplets, $N\gtrsim10^5$,  initially  appear in the 
region of the drop facing the electron source\cite{Lai19,Ell07}
and quickly  move due to  Coulomb repulsion,  until reaching their   lowest energy configuration 
on the droplet surface. The ionization of
pure helium nanodroplets has been discussed in Ref. \onlinecite{Mau18}; 
see also Ref. \onlinecite{Tig07}.
The case of negatively charged helium droplets is completely different 
as, because of Pauli's principle, bubbles of $\sim 19$ \AA{} radius void of helium atoms are formed 
around the negative charge\cite{Elo02,Gra06,Mar08} and are quickly ejected out of the droplet.
  
Highly charged helium nanodroplets could also be produced
by doping them  with neutral atomic impurities  
 and subsequently ionizing them. 
One interesting possibility is to start from droplets doped with several 
alkali (Ak) atoms, which are known to reside in dimples on the droplet 
surface.\cite{Anc95,Sti96,Bue07} The number of Ak dopants
can be adjusted by choosing the appropriate doping conditions.\cite{Toe04}
Once equilibrated, they can then be  photo-ionized, yielding  a 
full-extrinsic multiply charged helium drop.
The heavier Ca, Sr and Ba alkaline earth atoms have also been shown to reside at the droplet surface,
in a deeper dimple than that of Ak atoms.\cite{Her07}
They shoud thus behave as Ak atoms in the production of  
multiply charged helium droplets.
It must be noticed that alkali or alkali-earth atoms 
tend to form dimers or even larger clusters on the droplet 
surface.\cite{Von02,Sch04,Tig07,Sta10}
This can in principle be avoided by using large droplets and appropriate doping conditions (low pressure). 
As an alternative to the  photo-ionization of  multiple  dopants, one might  consider 
multiply charging a pure helium droplet before the  pickup   chamber:\cite{Fei22,Tie20} 
 doping  will then preferentially 
occur at the charged sites on the droplet surface and dopant ionization will proceed by charge transfer from He$^+_3$ ions.

Whereas the equilibrium position of one single Ak$^+$ ion 
in a He$_N$ droplet is at the center,\cite{Lea14}
multiple ions  --if they can coexist-- are expected to be evenly distributed close to the droplet surface.
This follows from the balance between  Ak$^+$-Ak$^+$ Coulomb repulsion and droplet-Ak$^+$ attraction. 
Beyond a critical $n$ value, the (Ak$^+)_n$He$_N$ drop will no longer be stable and Ak$^+$He$_M$ microclusters 
with $M \ll N$ will be ejected. 
 The same is expected for the heavier alkaline earth cations.
Charge location for intrinsic multiply charged droplets has recently 
been determined by X-ray coherent diffractive imaging with Xe as a contrast agent.
 Xe atoms clustered around the charges, revealing their distribution on the drop surface.\cite{Fei22}
  Hence, multiply charged helium droplets   constitute a realistic model system  for 
 Thomson's problem, i.e.,   for determining the minimum Coulomb energy configuration of 
$n$ charges constrained to remain on the surface of a sphere.\cite{Erb91,Mor96}
So far, only negative charges under the form of multielectron bubbles in superfuid helium\cite{Tem07,Guo08,Yad21} have been studied in this context.
Since electron-helium interaction is repulsive, whereas the interaction of helium with a positive charge is strongly attractive, 
significant differences can be expected in the surface structure and stability of these multiply charged spherical objects.

The liquid drop model (LDM), successfully used   to study 
nuclear fission,\cite{Boh98} has also been applied to address
Coulomb fragmentation of charged Van der Waals\cite{Ech88} and simple metal clusters,\cite{Gar95,Nah97} 
assuming that the process is triggered by Rayleigh instability: \cite{Ray82}
if the charge density reaches a threshold value, the drop 
breaks up into two daughter charged droplets of similar size due to  Coulomb repulsion. 
To compute the Coulomb contribution to the total energy, the LDM usually assumes that the
elementary charges are distributed uniformly within the system.
Therefore,   its application to the present problem is little justified,\cite{Mah07}  since
 charges are localized and Coulomb explosion proceeds by  ejection of  charged helium microclusters and not by droplet fission.\cite{Lai19}
In addition, these microclusters  (Atkins' snowballs) are highly structured around the charge. 
We propose here a  different  model based on the He-DFT approach\cite{Dal95,Bar06,Anc17,dft-guide} that circumvents 
the use of the LDM in conjunction with Rayleigh's stability criterion for
charged liquid drops. It also has the additional  advantage of realistically describing the snowball structure.  
  
Determining the minimum electrostatic energy configuration of $n$ equal 
point charges constrained to the surface of a unit sphere 
(Thomson's problem) is a classical problem that has no 
mathematically exact solutions except for a few cases:
antipodal points across a diameter  for $n=2$; an equilateral triangle at 
the equator for $n=3$; a regular tetrahedron for $n=4$; a triangular 
bipyramid  for $n=5$; a regular octahedron for $n=6$; and a regular icosahedron for $n=12$. 
For the general case,  global minimization of the electrostatic energy has to be performed numerically. 
The solutions for different $n$ values can be found in, e.g., Refs. \onlinecite{Erb91,Mor96}. 

In this work we address  the structure and energetics of extrinsic multiply charged helium droplets.
We design a model based on solvation energy determination within the He-DFT approach,  to which Coulomb energy is added.
The model is then applied to Na$^+$ and Ca$^+$ ions as case studies. 
Na or Ca atoms are assumed to be initially far enough  apart on the droplet surface that no dimer nor larger cluster is formed prior to ionization.

This manuscript is organized as follows. 
In Sec.~II we  introduce  the solvation plus Coulomb model proposed in this work. 
The  application of the  He-DFT approach to a doped planar helium surface
(as a model for the surface of a very large droplet) 
is described in Sec.~III. Results are  presented and  discussed in Sec.~IV,  
and a summary of the results is presented in Sec.~VI together with some concluding remarks.
Finally, the Appendix describes the determination of the  {\it ab initio} Ca$^+$-He potential 
used in this work. 

\section{The solvation plus Coulomb approach}

A full He-TDDFT simulation of a multiply charged droplet is computationally not feasible because of the large size required for droplets to 
host even a small number $n$ of charges.\cite{Lai19} 
To circumvent this limitation, we have  designed a model 
based on a crucial input:  the {\it solvation
energy} of the ion, which we calculate using 
 the He-DFT method. This model
is described in the following.
 
The condition for the existence of a stable configuration of $n$ charges \Ap\ in a spherical helium 
 droplet of radius $R$,  \nApHe n R,
is that its formation energy must be negative, i.e., 
the total energy of the ion-doped helium drop, $E[\nApHe n R]$, must be  lower than that 
of the reference system formed by He$_R$ and the $n$ charges at infinite distance
from the droplet and from each other, $E[\He R] + n E_{micro}$
\begin{equation}
\label{eq1new}
E[\nApHe n R] -  E[\He R] - n E_{micro} \le 0
\end{equation}
Note that the separated charges of reference can be bare \Ap\ or 
embedded  in a microcluster of helium atoms
to be defined below, 
hence their energy is denoted by $E_{micro}$. 
 
Assuming that  all the charges are sitting at a distance $d$ from the droplet surface, 
defined as the ``sharp density'' surface (where the helium atom density equals 
$\rho_0/2$ with $\rho_0=0.0218$ \AA$^{-3}$  the bulk liquid helium density),
the total energy can be expressed as
\begin{equation}
\label{eq2new}
E[\nApHe n R] = E_d[\nApHe n R]  + E_{Coul}[n\Ap](R+d) \; ,
\end{equation}
where $E[\nApHe n R]$ has been split into 
the interaction energy of the ions with the helium droplet, $E_d[\nApHe n R]$, 
and the Coulomb repulsion energy between ions, $E_{Coul} [n \Ap](R+d)$.
It is the attractive helium--ion  interaction that makes it possible for the multicharged drop to be stable.

Let us define the solvation energy $S_d[\nApHe n R]$ of $n$ positive ions in a helium droplet of radius $R$ as
\begin{equation}
\label{eq3new}
S_d[\nApHe n R]=E_d[\nApHe n R] - E[He_R] - n E_{micro}
\end{equation}
If the droplet is large enough that the interaction energy of one cation with the droplet is not affected by the presence of the others,
the solvation energy of $n\Ap$ simplifies to
\begin{equation}
S_d[\nApHe n R] \simeq n S_d[\ApHe R] 
\label{eq4new}
\end{equation}
where the zero for energies is taken as $E[\He R] $ for convenience.
The condition for stability (\ref{eq1new}) can then be expressed as
\begin{equation}
n S_d[\ApHe R] + E_{Coul}[n\Ap](R+d) \leq 0 \label{eq5new}
\end{equation}
For $n$ cations and a droplet of radius $R$, if a distance $d$ can be found 
such that inequality (\ref{eq5new}) is fulfilled, the multicharged 
droplet will be stable, i.e., will form a bound system.
Note that  $d$ can be positive or negative.

It is worth recalling that in order  to obtain the 
stability condition (\ref{eq5new}), we  
have implicitly assumed that the droplet is large enough 
and that all charges are at the same distance 
from the sharp density surface. In particular, 
the charge distribution is assumed to be the solution of Thomson's model on a sphere with radius $R+d$.
We will check in the following that this is a very reasonable approximation.

Calculating $S_d(\ApHe R)$ for large $R$ values can be very time consuming, especially if this has to
be done for many different $R$ values and locations of the ions  with respect to the droplet surface.
On the other hand, for big enough drops, curvature effects are small and can be 
safely neglected.  $S_d(\ApHe R)$   can 
then be approximated by replacing the drop with  a planar helium surface (PHeS). 
In this case,
the grand potential $\Omega= E- \mu N$ has to be considered rather than the energy, yielding
\begin{eqnarray}
S_d(\ApHe R) &\simeq& S_d(\Ap@PHeS)
\nonumber
\\
&=&  \Omega_d[\Ap@PHeS] -  \Omega[PHeS] - \Omega_{micro} 
\label{eq5}
\end{eqnarray}
 Hence  the problem  is   reduced to determining the grand potential 
$\Omega_d[\Ap@PHeS]$ for the ion at different distances $d$ 
from the sharp density surface of the PHeS,  the grand 
potential of the pristine PHeS, and that of the microcluster structure, all evaluated 
at the chemical potential of  bulk liquid helium at zero temperature and pressure.

The grand potentials are expressed as
\begin{eqnarray}
\Omega_d[\Ap@PHeS]
 &=& \int d \mathbf{r} [{\cal E}_d(\Ap@PHeS) - \mu \rho_d(\Ap@PHeS)] 
\nonumber
\\
\Omega[PHeS]&=& \int d \mathbf{r} [{\cal E}(PHeS) - \mu \rho(PHeS)]  
\label{eq6}\\
 \Omega_{micro}&=&\int d \mathbf{r} [{\cal E}(\Ap \text{He}_n)- \mu \rho(\Ap \text{He}_n)]
 \; , 
 \nonumber
\end{eqnarray}
where
 ${\cal E}$ is the energy density, 
$\mu$ is  the chemical potential of bulk liquid helium at zero temperature 
and pressure (-7.145 K for the functional we are using\cite{Anc05}), 
  $\rho$ is the helium atom density,
  and $d$ is the distance of 
the ion to the sharp density surface of the PHeS.
Finally,  $\Omega_{micro}$ is the grand potential of the microcluster, which is then defined as the cluster of $M$ heliums around the cation,
 with $M$ determined such that the chemical  potential is equal to that of bulk liquid helium.

Since we are assuming that the charges are distributed according to the 
solutions of Thomson's model, the
Coulomb energy can be written as\cite{Erb91} 
\begin{equation}
E_{Coul}[nNa^+] = \frac{e^2}{4\pi\epsilon_0}\frac{\xi_n}{R+d} \; ,
\label{eq7}
\end{equation}
where $e$ is the elementary charge and $\epsilon_0$ the vacuum electric permittivity.
The parameter $\xi_n$ depends on the number of charges $n$, for instance: 
$\xi_2 =1/2$, $\xi_3= \sqrt{3}$, $\xi_4=3\sqrt{6}/2$.  $\xi_n$ values up to $n=65$
are tabulated in Ref. \onlinecite{Erb91}; $\xi_n$ values for larger $n$ can be found in Ref. \onlinecite{Mor96}. 

The stability of a multicharged droplet is therefore determined by the 
function
\begin{equation}
U(r) = n S_d[\Ap@PHeS] + \frac{e^2}{4\pi\epsilon_0}\frac{\xi_n}{R+d} \; ,
\label{eq8}
\end{equation}
where $r=R+d$ is the distance of the charges  to   the center of the droplet
(recall that $d$ may be negative or positive).
$U(r)$ is the  
formation energy --referred to the infinitely separated droplet and ion microclusters-- of a  
droplet of radius $R$ hosting $n$ charges at a distance $d$ from its sharp density surface.
A distribution of charges with a common distance $r$ from the center of the droplet
will be energetically stable as long as $U(r)<0$.
 Therefore, the stability limit for a droplet containing $n$ charges corresponds to the  {\it smallest} value $R_0$ of its radius for 
 which  $U(r)=0$.  This yields
\begin{equation}
R_0(n) = min _{\{d\}} \left\{\frac{e^2}{4\pi\epsilon_0} \frac{\xi_n}{n |S_d[\Ap@PHeS]|} -d\right\} 
\label{eq9}
\end{equation}
For $R>R_0(n)$, 
solvation energy will dominate over Coulomb repulsion  so that  the energy of 
the system will be negative and the system will be stable. For $R<R_0$, 
 electrostatic repulsion will dominate over solvation energy
 and  the energy of the system will be positive: 
 the system is then expected  to be unstable against Coulomb explosion. 
 It is thus tempting to interpret $R_0(n)$  as the critical size observed in experiments.\cite{Lai19,Fei22}
 As will be seen  in the following, the existence of energy
barriers preventing  Coulomb dissociation of the charged system  drastically changes this simple perspective.

An essential ingredient in Eqs.~(\ref{eq8}-\ref{eq9})
is the solvation energy for one single ion at a distance $d$ from the planar helium surface, 
$S_d[\Ap@PHeS]$. It is obtained by carrying out 
 constrained He-DFT calculations similar to those employed 
for helium drops.\cite{Her08,Lea16,Cop17}  In the next section we describe in some detail the procedure 
 before discussing the consequences of Eq. (\ref{eq8}).

\section{He-DFT description of the doped helium planar surface}

Helium density functional theory (He-DFT), in its static and in its 
time-dependent versions, has proven to be
a very powerful tool to study the properties  
of superfluid $^4$He samples. It is a phenomenological approach
which constitutes a good compromise between accuracy and feasibility. 
A detailed description of the method, mainly applied to the study of 
helium droplets, can be found in Refs. \onlinecite{Bar06,Anc17,dft-guide}.
Within the He-DFT approach, the finite range of 
the helium-helium van der Waals interaction is explicitly  
incorporated, and surface properties of liquid helium, in
particular its surface tension, are well reproduced.
We have carried out calculations for Na$^+$ and Ca$^+$ ions doping a 
PHeS  within He-DFT in order to obtain $\Omega_d[\text{Na}^+@PHeS]$
and $\Omega_d[\text{Ca}^+@PHeS]$, as defined in the previous Section. 
Here we outline the method we have used for the calculations.

In the static calculations carried out in this work, 
the ions are described classically, hence their influence on the helium samples is described as  an external field.\cite{Anc17} 
Within the He-DFT approach at zero temperature, the energy of a $N$-atom helium sample $^4$He$_N$ 
doped with an \Ap\   ion located at $\mathbf{r}_{\Ap}$ is written as a functional of the $^4$He atom 
density $\rho({\mathbf r})$  as
\begin{eqnarray}  
&&E[\Psi, \mathbf{r}_{\Ap}] = \int d \mathbf{r} \, \frac{\hbar^2}{2m_{\rm He}}|\nabla \Psi|^2 +  \int d \mathbf{r} \, {\cal E}_c(\rho)  
\nonumber
\\
&&+ \int d \mathbf{r} \,  V_{\rm He-\Ap}(|\mathbf{r}-\mathbf{r}_{\Ap}|) \, \rho(\mathbf{r})  \; ,
\label{eq10}
\end{eqnarray}
where the first term is the  kinetic energy of the superfluid, 
$m_{\rm He}$ is the mass of the $^4$He atom, and
$\Psi({\mathbf r})$ is the effective wave function (or order
parameter) of the superfluid such that 
$\rho({\mathbf r})=|\Psi({\mathbf r})|^2$ with $\int d{\bf r}|\Psi({\bf r})|^2 = N$. 
The functional ${\cal E}_c(\rho)$ we have used contains the
He-He interaction term within the Hartree approximation and additional terms
describing non-local correlation effects.\cite{Anc05}
It is a modification of the Orsay-Trento functional\cite{Dal95} which makes it stable even in the presence of 
very attractive impurities. The interaction  
of one single helium atom
with the Na$^+$ ion has been taken from Koutselos et al.,\cite{Kou90} while 
in the case of Ca$^+$ it has been calculated 
{\textit{ab initio} and fitted to an analytical form  as indicated in the Appendix.

As already stated above, for the PHeS the grand potential  
$\Omega = E - \mu N$ has to be minimized rather than the energy, with $\mu$ the 
helium
chemical potential at zero temperature and pressure (-7.145 K). 
This guarantees the correct asymptotic density  in the 
bulk of the liquid helium.
The equilibrium density in the presence of  an \Ap\  ion is obtained by solving the 
Euler-Lagrange (EL) equation resulting from functional variation 
of the grand potential with $E$ given by Eq.~(\ref{eq10}), namely
\begin{equation}
{\cal H}[\rho] \,\Psi    = \mu \Psi \, ,
\label{eq11}
\end{equation}
where $\cal{H}$ is the DFT Hamiltonian
\begin{equation}
{\cal H}=  -\frac{\hbar^2}{2m_{\rm He}}\nabla^2 +
  \frac{\delta {\cal E}_c}{\delta \rho(\mathbf{r})} + V_{\rm He-\Ap}(|\mathbf{r}-\mathbf{r}_{\Ap}|)
\label{eq12}
\end{equation}

The effective wave function  $\Psi(\mathbf{r})$ is determined at the nodes of a three-dimension Cartesian grid of  0.3 \AA{} space step
inside a large calculation box. The EL equation has been solved by a relaxation (imaginary time $\tau$) 
method using a modified version of the $^4$He-DFT BCN-TLS 
computing package\cite{Pi17} (see  Refs.~\onlinecite{Anc17,dft-guide} 
and references therein for additional details) adapted to the PHeS geometry. 
Schematically, the relaxation method proceeds as
$\Psi(\tau + \delta \tau) = \Psi(\tau) + \delta \Psi(\tau)$, with
\begin{equation}
\delta \Psi(\tau)= - \frac{\delta \tau({\cal H}-\mu)}{1-(\langle {\cal H}\rangle- \mu)\delta \tau}\, \Psi(\tau) \; ,
\label{eq13}
\end{equation}
where $\langle {\cal H}\rangle= \langle  \Psi(\tau)|{\cal H}| \Psi(\tau)\rangle$.
We  have imposed specular symmetry at the surface
of the box in all three Cartesian coordinates,
i.e., continuity of the helium density  and cancellation of its first derivative. 
This allows us to use  cosine  Fast Fourier Transform\cite{Fri05}
to efficiently compute the convolutions needed to obtain the DFT
mean field ${\cal H}[\rho]$. The differential operators 
in ${\cal H}[\rho]$ are approximated by 13-point formulas.
 
In order to fix the planar helium surface at an arbitrary location such that $\langle z \rangle = z_0$, 
where $z$ is the 
Cartesian coordinate in the direction perpendicular to the helium surface and
\begin{equation}
\langle z \rangle = 
\frac{\int d \mathbf{r} \,z \,\rho(\mathbf{r})}{\int d \mathbf{r} \, \rho(\mathbf{r})}  \; ,
\label{eq14}
\end{equation}
we have  added a constraint term to the energy $E$  in Eq.~(\ref{eq10}).
This constraint is equal to $\lambda_c(\langle z\rangle -z_0)^2/2$,\cite{Her08,Lea16,Cop17} where 
$\lambda_c$ is an arbitrary constant large enough to ensure that 
upon minimization,  the imposed value $\langle z \rangle = z_0$   is obtained.
This  constrained minimization  is needed to avoid the homogeneous liquid solution, which also has $\mu = -7.145$ K. 
We have taken $\lambda_c = 5 \times 10^5$ K \AA$^{-2}$, which 
ensures an optimal convergence rate.
The desired $z_0$ values are then obtained to within  0.1\% accuracy. 

The overall procesure is as follows:  \textit{(i)} Starting from a reasonable density guess, carry out a 
constrained minimization of the grand potential for a  pristine PHeS with  a $z_0$ value well inside the calculation box. 
This 
 has to be done only  once, and it  yields  $\Omega[PHeS]$ 
for the pristine helium surface. It can also be used
to determine the surface tension of the helium surface, see next section. 
\textit{(ii)} Repeat the procedure including the \Ap\  ion at 
different $z_{\Ap}$ positions, and  minimizing the grand potential  with the 
same constraint $\langle z \rangle = z_0$ as before.  \textit{(iii)} 
In the result of step \textit{(ii)}, determine the position $z_{surf}$ of the sharp density surface at the box limits in the $x$ or $y$ direction (i.e., as far away 
from the ion as possible), and hence the distance $d=z_{\Ap}-z_{surf}$ of the ion from it. 

\section{Results}

\begin{figure}[!]
\centerline{\includegraphics[width=1.0\linewidth,clip]{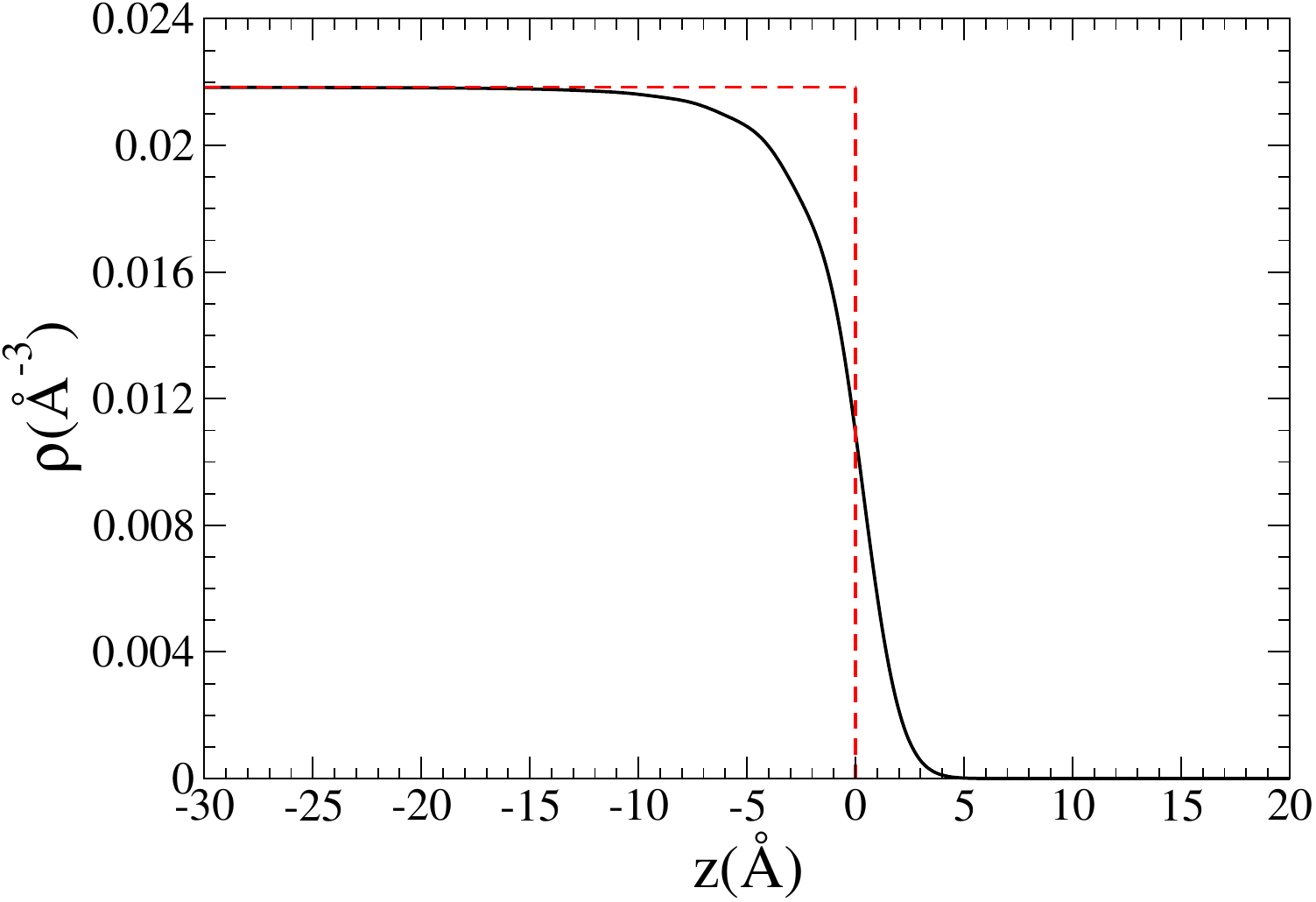}}
\caption{
 Helium density profile  in the direction $z$
perpendicular to its planar surface. 
The origin of the $z$-axis is arbitrarily set at the location of the sharp density surface (red dashed line).
}
\label{fig1}
\end{figure}

The density profile of the pristine planar helium surface at equilibrium is plotted in Fig.~\ref{fig1} along the $z$ direction, perpendicular to the surface. 
As mentioned above, 
 the surface tension $\gamma$ of the liquid can be evaluated from the equilibrium density.
It is given by
\begin{equation}
\gamma =\frac{1}{\cal{S}}\int d \mathbf{r} [{\cal E}(He_{PHeS}) -  \mu \rho(He_{PHeS})] \, ,
\label{eq15}
\end{equation}
where $\cal{S}$ is the area of the planar surface. 
The resulting value is $\gamma=0.278$ K \AA$^{-2}$ with the functional used in this work.\cite{Anc05} 
It is in good agreement with the experimental value for $^4$He, 0.274 K \AA$^{-2}$.

Two-dimensional plots of the density for different distances $d$ of the cation to the planar helium surface 
are displayed in Fig.~\ref{fig2} for Na$^+$ and in Fig.~\ref{fig3} for Ca$^+$. 
The plane is perpendicular to the surface and contains the ion.
Both  figures reveal  high-density structures around the ions. 
In addition, density lumps can be seen around the Ca$^+$ ion due to spontaneous symmetry breaking.
This is not the case around the Na$^+$  where spherical symmetry is conserved.
 Although the
Na$^+$-He potential well is more attractive than the Ca$^+$-He  one, 
it is narrower (see Fig.~\ref{fig4}), which prevents the density lumps from developing.

\begin{figure}[!]
\centerline{\includegraphics[width=1.0\linewidth,clip]{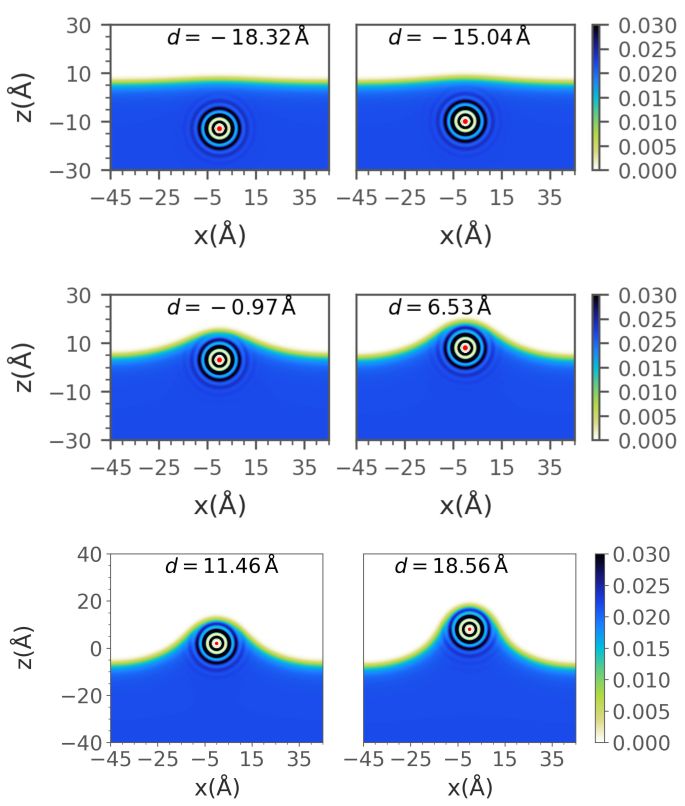}}
\caption{
 Two-dimensional plots  
of the density corresponding to a planar He surface hosting a Na$^+$ ion at a distance  $d$ 
from the sharp density surface. The color bar shows the density in units of \AA$^{-3}$.
 From left to right and from top to bottom: $d= -18.32$, $-15.04$, $-0.97$, $6.53$, $11.46$, and $18.56$~\AA.
Densities are shown in a plane perpendicular to the helium surface passing through 
the ion position.
}
\label{fig2}
\end{figure}

\begin{figure}[!]
\centerline{\includegraphics[width=1.0\linewidth,clip]{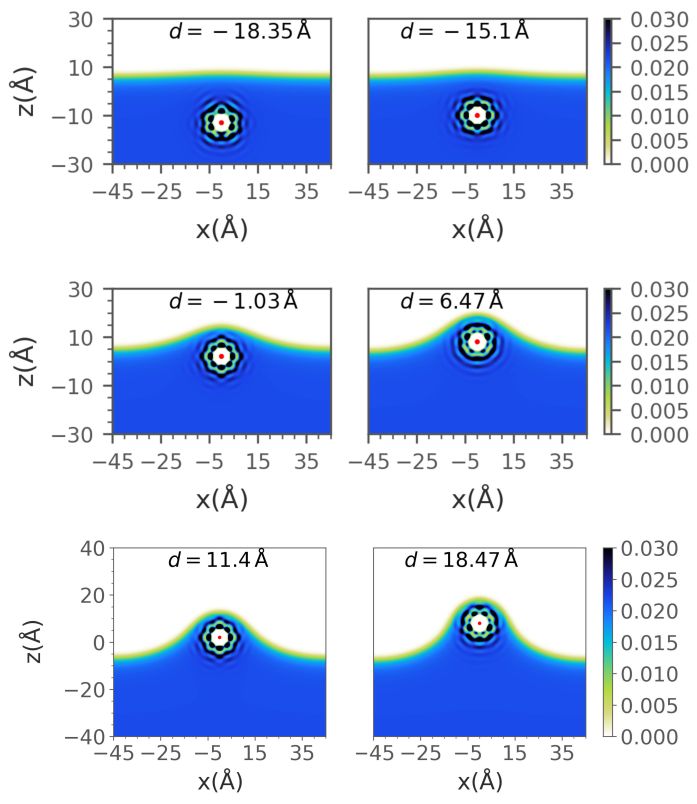}}
\caption{
  Two-dimensional plots  
of the density corresponding to a planar He surface hosting a Ca$^+$ ion at a distance $d$ 
from the sharp density surface. The color bar shows the density in units of \AA$^{-3}$.
 From left to right and from top to bottom: $d= -18.35$, $-15.10$, $-1.03$, $6.47$, $11.40$, and $18.47$~\AA.
Densities are shown in a plane perpendicular to the helium surface passing through 
the ion position.
}
\label{fig3}
\end{figure}

Another ingredient that plays an important role in the model is that of the microcluster, introduced in 
 Eq.~(\ref{eq1new}) and further defined following Eq.~(\ref{eq6}).
As shown in a number of mass spectrometry experiments, ions  produced by electron impact ionization of doped helium nanodroplets can be surrounded by
many helium atoms.\cite{Gon20}
The mass of the dopant as well as the ionization process and the settings of the ion source have a strong influence
on the relative abundance of helium atoms around the dissociated ions. 
The situation described in the present simulations corresponds to the limit of very slow dissociation of the ions (``adiabatic'' limit):
the dissociating  ions are solvated inside microclusters such as the ones starting 
 to emerge out of the helium surface at the larger values of $d$  in Figs \ref{fig2}-\ref{fig3}.
As long as these microclusters are still attached to the rest of the helium, even by a tiny density bridge,
their chemical potential must be equal to that of bulk liquid, which determines the cluster size and grand potential $\Omega_{micro}$. 
The other extreme model would be that of infinitely fast dissociation of the ions (``infinite order sudden'' approximation), where the ions would 
leave the droplet so fast that the surrounding helium could not follow and they would emerge as bare ions.
Although a case of such fast dissociation has already been observed, it was in the very specific case of sudden ionization of alkali dimers 
sitting on a droplet surface,\cite{Kri22,Kri23,Gar24b} where Coulomb repulsion was very strong and ``solvation'' was very weak.
Both these models, adiabatic and sudden, 
should be considered as opposite extremes of the 
same physical process, while the actual scenario as probed in experiments 
is likely in between. This should be considered when comparing our predictions 
with the experimental results, and could justify some quantitative discrepancies 
with our results, as will be discussed in the following.

In order to compute $\Omega_{micro}$, we have simulated the structure of the microcluster around the \Nap\ and the \Calp\ ions separated 
from the PHeS, with the chemical potential set at the value of the bulk liquid and the number of helium atoms (i.e., the integral of the helium density) 
to be determined. 
Figure \ref{fig5} shows the  density profile of these microclusters.
Notice the solid-like character of the first helium shell around Na$^+$, whereas 
 the one around Ca$^+$ looks more like  that of a structured liquid.
Both microclusters are fairly big: they contain a total of 105 helium atoms for \Nap\ and 93 for \Calp.
 As can be seen in Fig.~\ref{fig5}, they extend way beyond the  first solvation shell, 
which contains 14 helium atoms for \Nap\ (values in the 9-16 atoms range are found using different approaches, see table II of Ref. \onlinecite{Gar24}), 
and 27 for \Calp (close  
to the value of 25 found by Bartolomei et al. \cite{Bar21}). 
These large atom numbers are a likely consequence of the adiabatic approximation 
underlying the static constrained
calculations of  our  approach.\cite{Her08,Lea16,Cop17} In the real ejection dynamics, 
many of these atoms are stripped off the
microcluster.\cite{Alb23}  
\begin{figure}[!]
\centerline{\includegraphics[width=1.0\linewidth,clip]{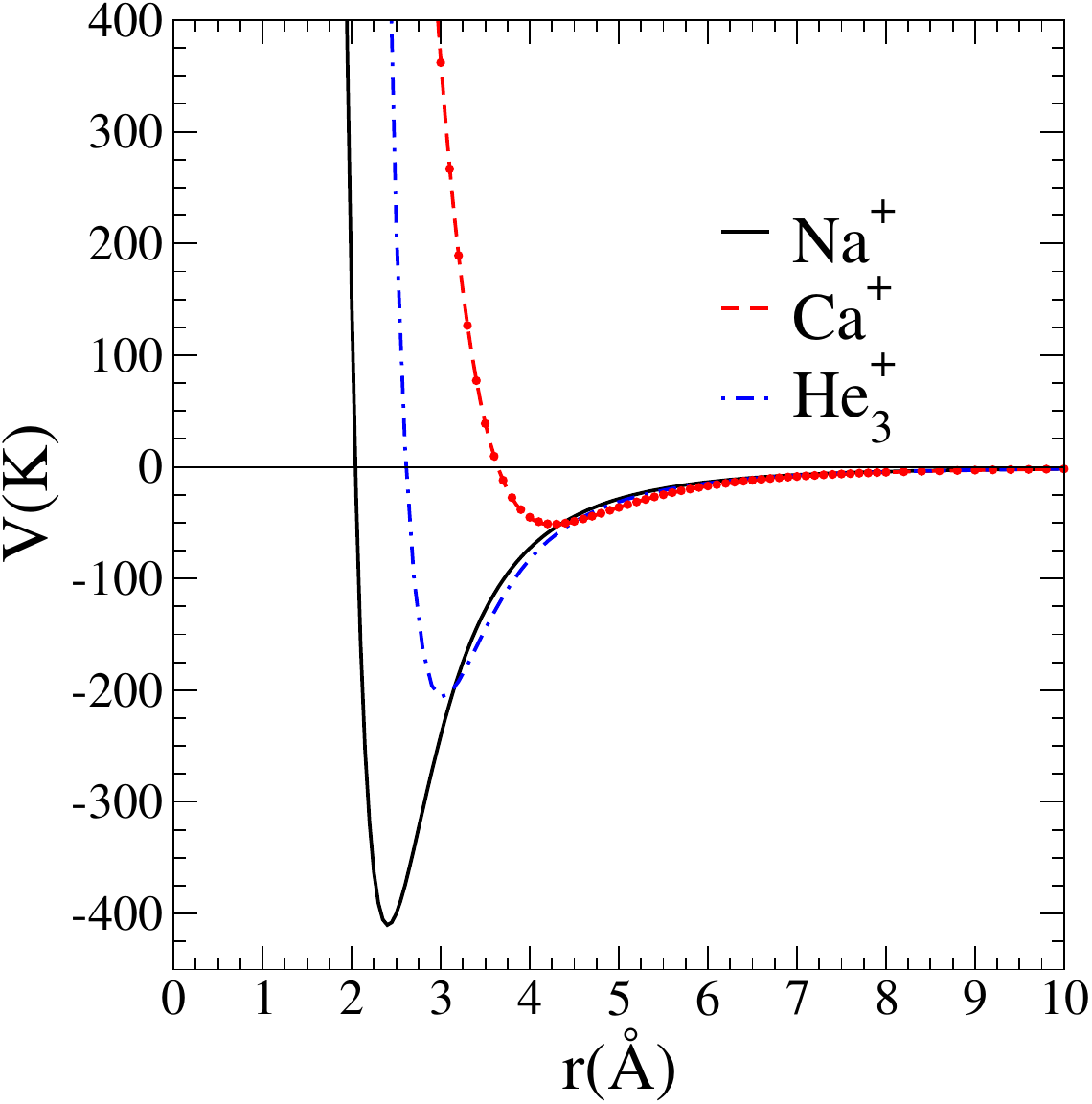}}
\caption{
Na$^+$-He pair potential,\cite{Kou90} spherically averaged He$_3^+$-He pair potential,\cite{Mat14} and Ca$^+$-He
pair potential calculated in this work.
}
\label{fig4}
\end{figure}

\begin{figure}[!]
\centerline{\includegraphics[width=1.0\linewidth,clip]{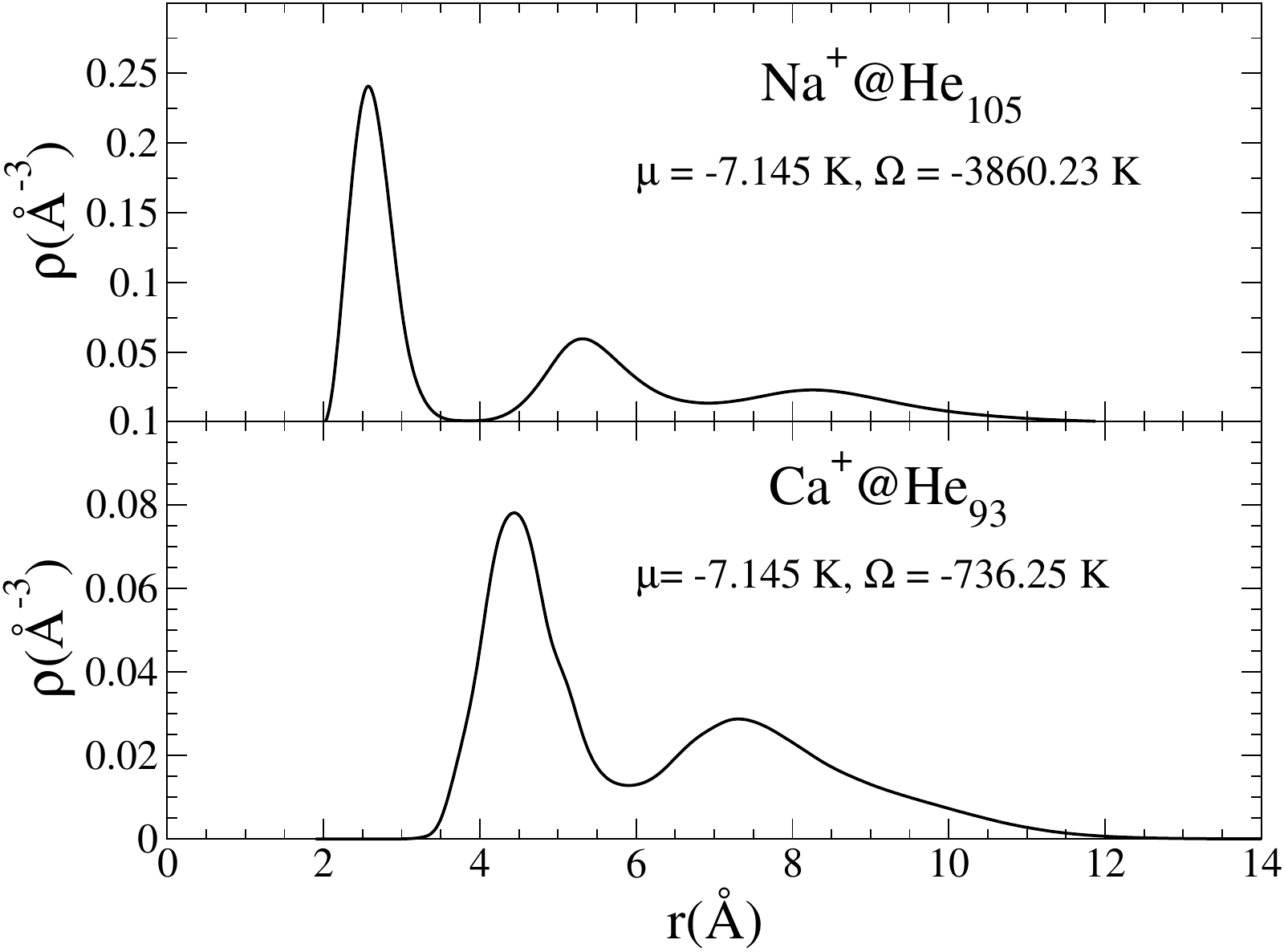}}
\caption{
Top panel: Density profile of the helium microcluster around a Na$^+$ ion 
at chemical potential $\mu=-7.145$ K. 
Bottom panel: Spherically averaged density profile of the helium 
microcluster around a Ca$^+$ ion at chemical potential $\mu=-7.145$ K.
The integrated number of helium atoms and the value of the corresponding grand potential ($\Omega$) are  also given 
 in the panels.
}
\label{fig5}
\end{figure}

We have determined the solvation energies  $S_d[\Nap@PHeS]$ and $S_d[\Calp@PHeS]$ 
 for a number of values  of the distance $d$ of the ion to the sharp density planar surface from Eq.~(\ref{eq5}).  
The results are displayed    in Fig.~\ref{fig6}.
In spite of the marked difference between the Na$^+$-He and Ca$^+$-He pair potentials, 
 which can be seen  in Fig. \ref{fig4}, the similarity between the solvation curves is remarkable.
Considering Eq.~(\ref{eq5}), 
we can conclude that once $\Omega_{micro}$ --the grand potential for the local structure around the ion--  
 is subtracted from the grand potential $\Omega_d[\Ap@PHeS]$ of the overall system, 
 the resulting solvation energy does not depend on the nature of the ion.
This reflects the fact that the specificity of ion-helium interaction is responsible for the structure of the microcluster,
and that beyond that structure, the ion-helium long distance interaction is species-independent, being that of a positive, elementary 
charge with helium (charge-induced dipole interaction).
For this reason, from now on  most of  the results will only be discussed   for the case of the Na$^+$ ion.

We  found it useful 
 for the remainder of this work  to fit the calculated values of
$S_d[\Nap@PHeS]$ and $S_d[\Calp@PHeS]$ to a simple analytic function.
A very good fit has been  obtained  using  the following expression
\begin{equation}
S_d[\Ap@PHeS]= \frac{S_0}{\left[1+exp\left(\frac{d-d_0}{\sigma}\right)\right]^{\nu}} 
\label{eq16}
\end{equation}
with $S_0= -797.641$ K, $\sigma=9.262$ \AA, $d_0=16.627$ \AA, and $\nu=1.234$
for \Nap; and $S_0= -784.998$ K, $\sigma=9.431$ \AA, $d_0=19.486$ \AA, and $\nu=1.533$ for \Calp. 
 As can be seen in Fig.~\ref{fig6}, some departure of the fit from the actual He-DFT values only appears for large negative $d$ values. 

Notice that the solvation curves shown in Fig.~\ref{fig6}
appear to flatten out 
 for ion positions deep inside helium.
On the contrary, mutual Coulomb repulsion increases strongly when ions get closer to the droplet center.
As a consequence, if the charges are initially produced deep inside the droplet, they will move towards the surface until Coulomb repulsion 
is compensated by solvation energy.
 
\begin{figure}[!]
\centerline{\includegraphics[width=1.0\linewidth,clip]{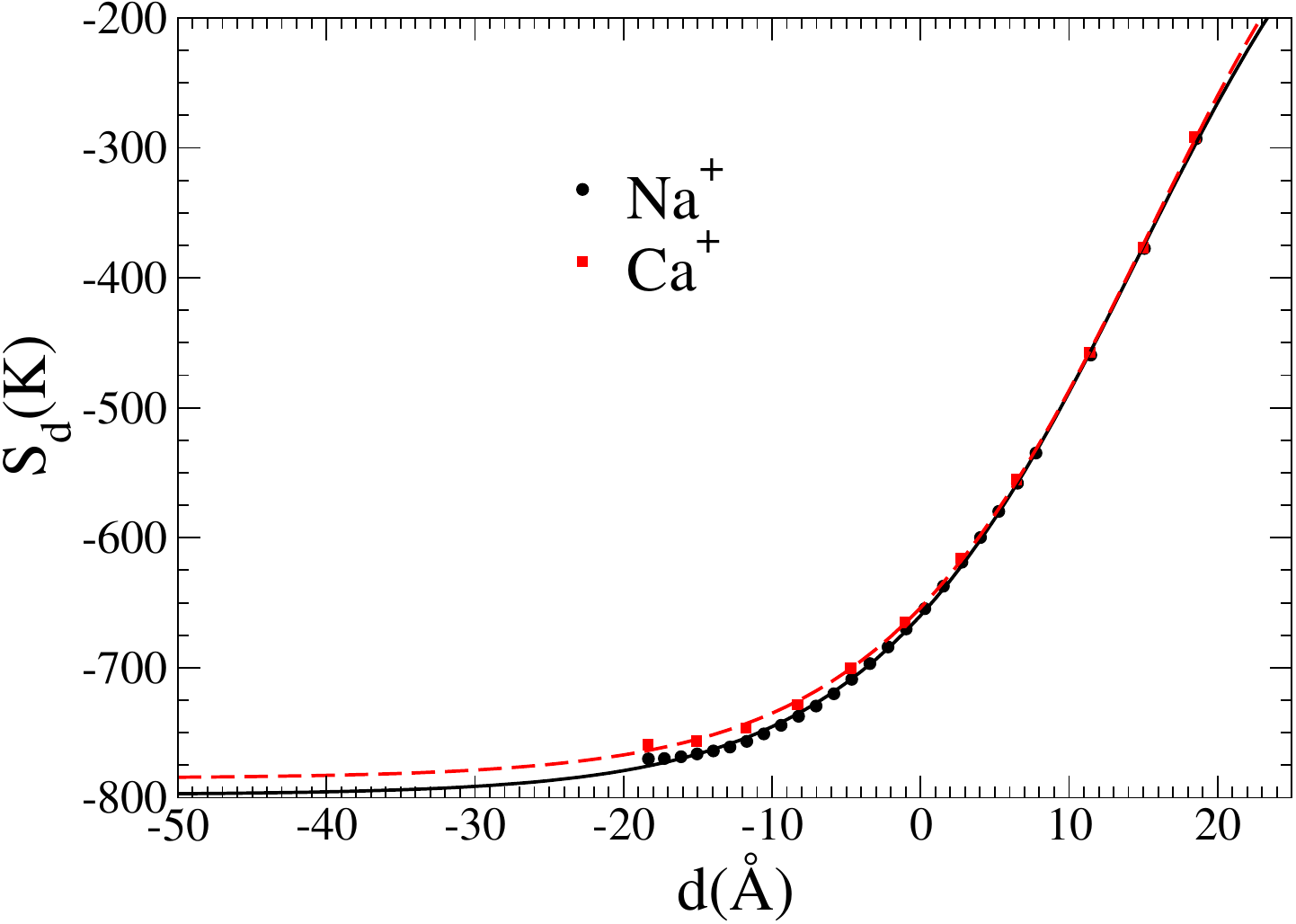}}
\caption{
He-DFT solvation energy of the ion as a function of its distance to the sharp density planar surface: $S_d[Na^+@PHeS]$ (black dots) and 
$S_d[Ca^+@PHeS]$ (red dots), and fit by Eq.~(\ref{eq16}): Na$^+$ (solid, black line) and Ca$^+$ (dashed, red line).
}
\label{fig6}
\end{figure}

Once $S_d[\Nap@PHeS]$ has been calculated, it is straightforward to determine 
the radius of the smallest droplet hosting an equilibrium distribution of
$n$ positive charges on its surface, $R_0(n)$, using Eq. (\ref{eq9}).
Figure \ref{fig7} shows the results
obtained for $n=2-10$, 15 and 20. 
The behavior is the same for both cations. It   
reflects the fact that the minimum Coulomb energy
of $n$ charges on a sphere of radius $R$ (Thomson's model) can be fitted by the empirical formula\cite{Erb91}
\begin{equation}
E_{Coul}= \frac{e^2}{4\pi\epsilon_0} \frac{1 }{R} \left[ \frac{n^2}{2}- 0.551\, n^{3/2} \right]
\label{eq17}
\end{equation}
and therefore 
\begin{equation}
R _0(n) \sim  \frac{e^2}{4\pi\epsilon_0}  \frac{1}{|S_d[\Ap@PHeS]|}\left[ \frac{n}{2}- 0.551\, n^{1/2} \right]  \; .
\label{eq18}
\end{equation}

The minimum size for a droplet to contain $n$ charges has been found to scale as $n^{1/2}$ in experiments.\cite{Lai19,Fei22}  
This is different from the dependence obtained here for the stability threshold radius $R_0(n)$. The  resulting $R_0(n)$ curve looks 
almost linear in Fig.~\ref{fig7}, in part  because of the small range of $n$  values.

Other estimates  have been proposed for the minimum size of charged liquid droplets that ensures stability 
against Coulomb repulsion.   
Livshits and Lozovik have addressed the crystallization and melting of a system of 
charges in a helium drop\cite{Liv07} within a LDM 
 in which all charges are uniformly smeared 
over the spherical surface.
Applying Rayleigh's criterion for the stability of a charged liquid droplet  
 against  surface oscillations,\cite{Ray82} they obtain
for the critical radius [see Eq. (17) of Ref. \onlinecite{Liv07}]
\begin{equation}
R _c =\left[\frac{e^2}{4\pi\epsilon_0}\frac{ n^2}{16 \pi \epsilon \gamma} \right]^{1/3}   \; ,
\label{eq19}
\end{equation}
where $\epsilon = 1.05$ is the dielectric constant of liquid $^4$He. 
For the sake of comparison with our results, we take $\epsilon= 1$ and
obtain $R_c = 23 \, n^{2/3}$ \AA{}.
Hence the scaling for the critical radius in this surface LDM plus Rayleigh model is then $R_c \propto n^{2/3}$, at variance with experiment.

\begin{figure}[!]
\centerline{\includegraphics[width=1.0\linewidth,clip]{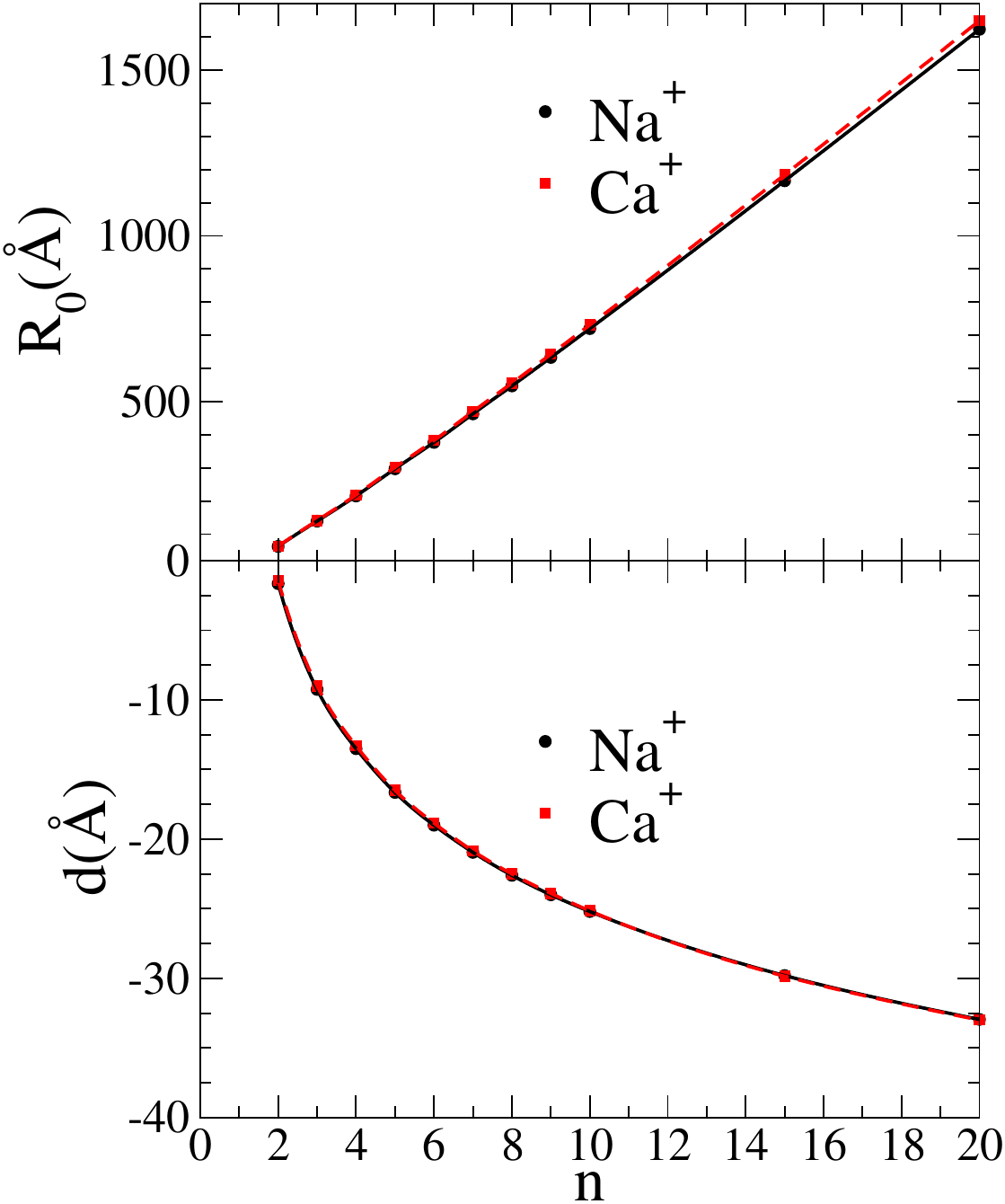}}
\caption{
Top panel: Limiting radius $R_0$ (\AA) for the stability of  helium droplets hosting $n$ Na$^+$ (black dots, black solid line) or Ca$^+$ 
(red dots, dashed red line) ions for $n=2-10$, 15 and 20.
Bottom panel: distance $d$ (\AA{}) of the ions to the sharp density surface of the droplet.
}
\label{fig7}
\end{figure}

We have checked an important assumption underlying the form of Eq. (\ref{eq8}),
namely that the equilibrium configurations of $n$ charges on the droplet surface 
 are  solutions of Thomson's model.
In particular, we wanted to check  whether  they shared
the same common distance $R+d$ from the droplet center.

In order to do that, we  applied  
a Steepest-Descent (SD) algorithm
to find the minimum
energy configuration of  $n$  ions subject to solvation plus Coulomb forces in a droplet of radius $R$.
In the iterative SD search algorithm, 
the positions $\mathbf{r}_i$ ($i=1,n$) of the $n$ ions were updated at each iteration as
\begin{equation}
\mathbf{r}_i^{ k+1}=\mathbf{r}_i^{ k}+\eta (\mathbf{F}_{solv,i}^ k+\mathbf{F}_{coul,i}^ k) \; ,
\label{eq20}
\end{equation}
where $\mathbf{F}_{coul,i}^k$ is the repulsive force acting on the $i$-th charge 
due to the other ions, and 
$\mathbf{F}_{solv,i}=-(dS_d/dr)(\mathbf{r} /r)$ is the solvation force
acting on the $i$-th charge due to the surrounding helium (attractive towards the 
droplet center), with $S_d$ taken from Eq.~(\ref{eq16}).
In addition,  $\eta$ is an arbitrary parameter to be adapted in order to speed up convergence:  if it is too small, it takes very long  
to move downhill along
a potential energy surface, and if it is too large one may have overshooting, i.e., 
the energy will increase from one step to the next. 
We have used $\eta=6 \times 10^{-5}$  \AA$^2$/K.
Relation (\ref{eq20})  was  iterated until the forces  cancelled out, i.e. an energy minimum  was  found,
 and hence  the  corresponding  ion-droplet equilibrium distances.
The initial ion positions were taken from the corresponding Thomson's structure with a small random displacement.
The more general displacements allowed by the SD dynamics  lift  the constraint of equal ion-center
distance, 
 even when  the initial configuration is 
  a solution  of the Thomson model,  allowing in principle for different radial displacements
(``buckling'') or changes in the ions equilibrium
positions on the droplet surface.

We  found   very small deviations with respect to the 
ideal Thomson's structures, and very small energy gains.
For instance, in the case of $n=20$ Na$^+$ ions on the surface of a helium nanodroplet of 
radius $R=250$ \AA{} (a value which 
  --barely-- guarantees   Coulomb stability of the system),    the average distance 
of the cations from the droplet center   is $\langle R+d \rangle=235.4$ \AA,  
  and  the maximum excursion from the above value is  $\sim 0.8$ \AA.
The gain in total energy with respect to the 
ideal solution is just 34 K over a total energy of $\sim 10^4$ K.
We have  recalculated  the  $R_0(n)$ curve using the SD scheme and
found results virtually indistinguishable from those shown in Fig. (\ref{fig7}).

Thomson's model solutions with $n$ charges are known to have a series
of ``magic'' numbers, where the lowest-energy structures are particularly stable:\cite{Mor96} $n=12, 32, 72, 122, \ldots$
Most of these structures are icosahedral.
We have verified that magic numbers are also present
for the structures found in our   SD minimization,  by computing
the second total energy  (Coulomb + solvation) difference
as a function of  cluster charge
\begin{equation} 
D_2[n]=E[n-1]-2E[n] +E[n+1] \; .
\label{eq21}
\end{equation}
Peaks in this quantity signal particularly stable structures.
Two magic numbers, $n=12$ and $32$, were found within  the range explored here, in agreement with Thomson's model.

Unexpectedly, during the SD optimization of ion positions, locally stable configurations were encountered for droplets with a  radius smaller than 
the stability value $R_0$, hence for which ions should be ejected.
These metastable configurations are a signal for the presence of  energy barriers preventing ion ejection. 
In order to clarify this important issue, we 
examined the  behavior of the energy $U(r)$, $r=R+d$, using  Eqs.~(\ref{eq8}) and (\ref{eq16}).

Figure \ref{fig8} shows $U(r)$
for $n=2$ and different $R$ values. 
One of the curves, corresponding  to $R=63.17$ \AA{}, has a minimum 
with value $U=0$, thus identifying $R_0$. Droplets with $R>R_0$ 
(e.g., $R=70$ \AA {} in the figure) are stable: they display 
a minimum with energy smaller than that of the reference system (i.e., a negative minimum).  
However, $U(r)$ also displays a positive minimum for many $R<R_0$ radii,
separated by energy barriers from the dissociated configuration.
These radii should correspond to {\it metastable} multicharged droplets, 
as their energy is higher than that of the 
reference system. 
These features are common to all the other $n$ values investigated here.

For smaller and smaller $R$ values, the barrier gradually decreases until disappearing: 
the charged droplet cannot be
metastable anymore and ions are directly  ejected off the droplet. 
For $n=2$ the critical radius at which the barrier disappears is $R=24.65$ \AA{}, as illustrated in Fig.~\ref{fig8}. 
We call this radius the ``explosion'' radius $R_{expl}$.
 Its value for Na$^+$ is plotted in Fig.~\ref{fig9} as $n$ vs $R_{expl}^2$ for the sake of comparison with experiment. 

 The barriers observed in  Fig.~\ref{fig8}  can be very high compared to thermal activation energy,
$E_{ther} = (3/2) k_B T$, with $k_B$  the Boltzmann constant.
The highest barrier is for $R=R_0$ ($\sim 700$~K for two Na$^+$ ions).
In helium droplets $T$ is very small, of the order of 0.4 K.\cite{Toe04}
In view of this, we  propose a sensible definition for the critical radius  $R_c(n)$  above which  a droplet with $n$ charges  is
stable at least up to the  millisecond  timescale of the experiments:\cite{Lai19,Fei22}  $R_c$ is the radius of the
droplet that displays an energy barrier equal to the thermal activation energy.  Given the energy scale in  Fig. \ref{fig8}, 
 $R_c \simeq R_{expl}$ in practice.
For instance, for $n=20$ we have found that the barrier for $R=219.5$ \AA{} (which we have taken as $R_{expl}$) 
is 0.06 K; for $R=219.65$ \AA, it is 0.67 K.

Also shown in Fig.~\ref{fig9}  are  the critical radii obtained by Livshits and Lozovik in the surface LDM plus Rayleigh instability 
approach,\cite{Liv07} Eq. (\ref{eq19})  with $\epsilon=1$, as well as the experimental $R_c$ 
values of Laimer \textit{et al.} parameterized as\cite{Lai19}
\begin{equation}
n= a + \zeta\, R_c^2 
\label{eq22}
\end{equation}
with $a= -0.544$ and $\zeta=2.6 \times 10^{-2}$ nm$^{-2}$.
For the sake of comparison with the experimental
results, we exceptionally give distances in nanometers in this figure and in the related discussion. 
As can be seen in Fig.~\ref{fig9}, $n$  depends  linearly on  $R_{expl}^2$ in the He-DFT approach, in agreement with the experimental results,
except for small $n$ values for which  curvature effects may be important.
The results obtained in the surface LDM plus Rayleigh instability approach exhibit a different behavior ($R_c^2\propto n^{4/3}$, 
see Eq.~(\ref{eq19}) and following text).

Figure \ref{fig10} shows the distance of the Na$^+$ cations to the surface of the droplet 
with $R=R_{expl}$. In all cases $d>0$, meaning that the  cations are above the droplet surface.
 For  $n\gtrsim 15$, ions are $\sim 15.5$ \AA{} away from the surface,  
 giving the droplet a ``virus-like'' appearance,
as shown for $n=20$  in the inset of the figure.
Notice the nearly triangular arrangement of the ions on the droplet surface.

The experimentally observed relation 
between $R_c$ and $n$ for pristine helium droplets\cite{Lai19} is
$n \sim \zeta \,R_c^2$, implying that the surface density of cations
at which  instability arises,
\begin{equation}
{\cal N}=n/(4 \pi R_c^2)=\zeta/(4 \pi) \; ,
\label{eq23}
\end{equation}
is constant. This seems reasonable: any instability
due to Coulomb repulsion must set in at the
same critical  charge surface density, irrespective
of the droplet size.
Using the experimental results of Ref. \onlinecite{Lai19},  one finds ${\cal N}_{expt}=2.07 \times 10^{-3}$  nm$^{-2}$.

One can estimate the average distance $\delta$
between neighboring ions on the droplet surface with critical radius $R_c$.
 Assuming that the $n$  charges are arranged in a triangular lattice  on the
surface of  the sphere  (this is the expected structure 
for large values of  $n$), the surface area per particle is
$\Sigma=\sqrt{3} \delta^2/2$,
$\delta$ being the nearest-neighbor distance. 
The total area $n \Sigma$ must be equal to $4 \pi R_c^2$, i.e.
$n \sqrt{3} \delta^2/2=4 \pi R_c^2$,
 and therefore $\delta=\sqrt{8 \pi/(\sqrt{3} \zeta})$. 
Using the experimental value of $\zeta$ one obtains $\delta= 23.6$ nm{}.
For $n\gtrsim 10$ our results for $R_{expl}$ also follow the law
$n = a' + \zeta' \,R_{expl}^2$ (see Fig. \ref{fig9}) with $a'=4.23$, $\zeta' =3.23 \times 10^{-2}$ nm$^{-2}$ and 
therefore $\delta=21.2$ nm and ${\cal N}= 2.57 \times 10^{-3}$ nm$^{-2}$, similar to the experimental one for He$_3^+$ ions.

\begin{figure}[!]
\centerline{\includegraphics[width=1.0\linewidth,clip]{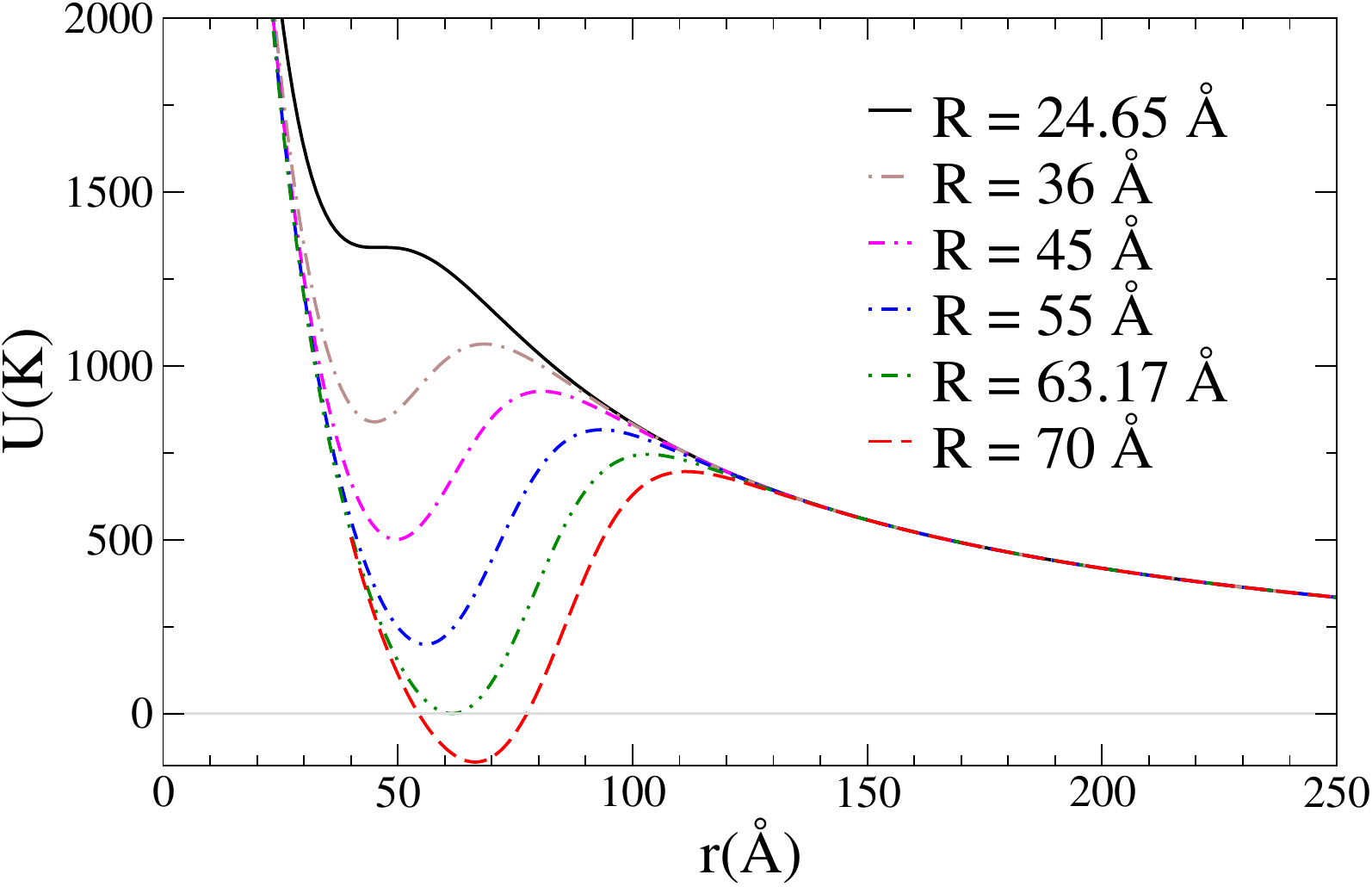}}
\caption{
Energy $U(r)$ of a droplet hosting 
two Na$^+$ ions as a function of the distance $r$ of the ions 
to the center of the droplet.
From top to bottom, the curves correspond to droplets with radius 
$R$=24.65 ($R_{expl}$), 36, 45, 55, 63.17 ($R_0$), and 70 \AA.
The origin for energies is that  of the reference (dissociated) system.
}
\label{fig8}
\end{figure}

\begin{figure}[!]
\centerline{\includegraphics[width=1.0\linewidth,clip]{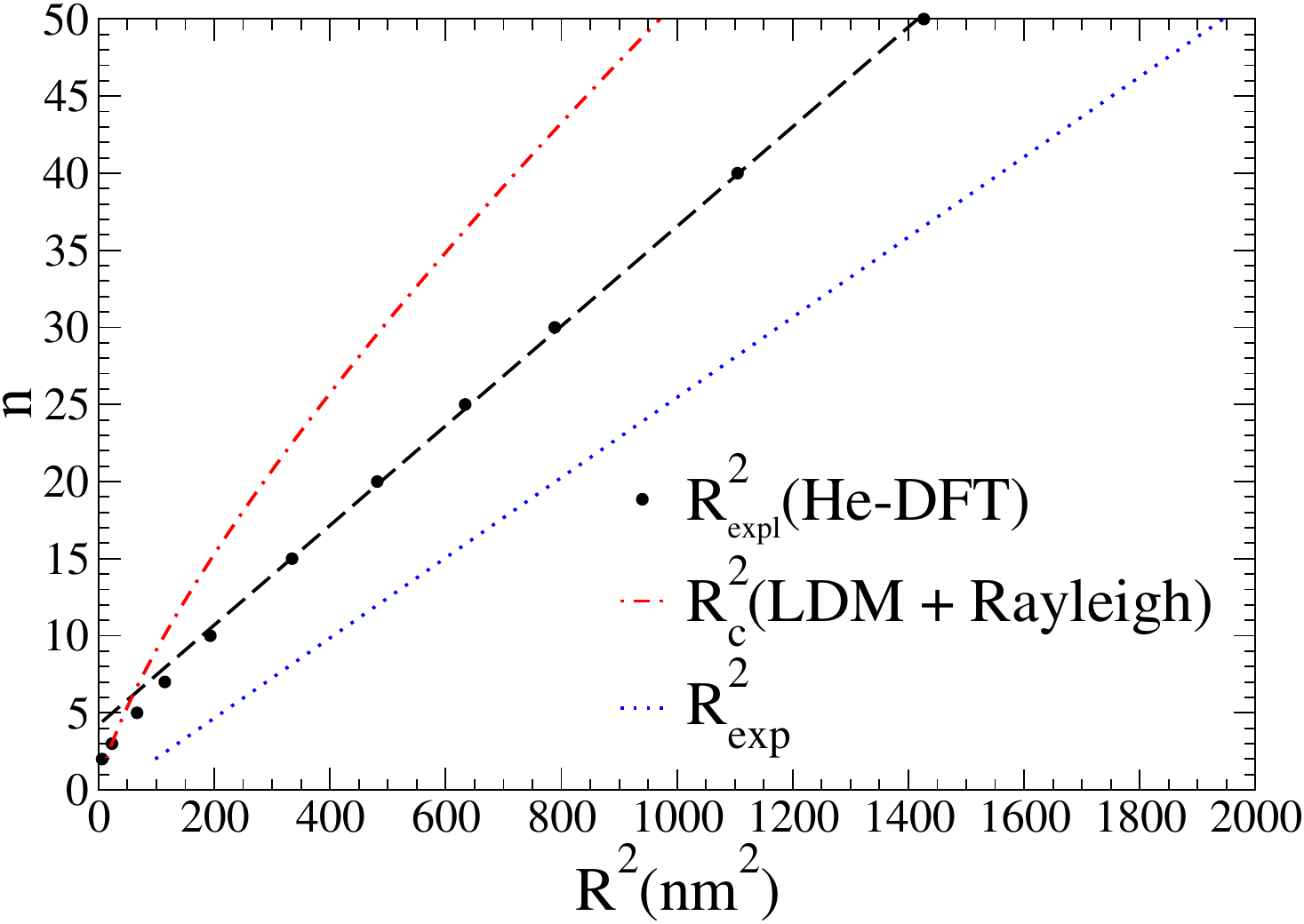}}
\caption{
Number of charges $n$ vs $R^2$ (in nm$^2$).
Black dots: $R^2=R^2_{expl}$ for Na$^+$ in the He-DFT approach,
 with $R_{expl}$ the minimum value below which multicharged droplets can no longer be even metastable (see text);
dashed line: linear fit ($n= a' + \zeta' R_{expl}^2$) to the results from $n=10$ to 50. 
Dotted line: experimental values\cite{Lai19}   $R^2=R_c^2$.
Dash-dotted line: $R^2=R_c^2$ in the LDM plus Rayleigh instability approach.\cite{Liv07} }
\label{fig9}
\end{figure}

\begin{figure}[!]
\centerline{\includegraphics[width=1.0\linewidth,clip]{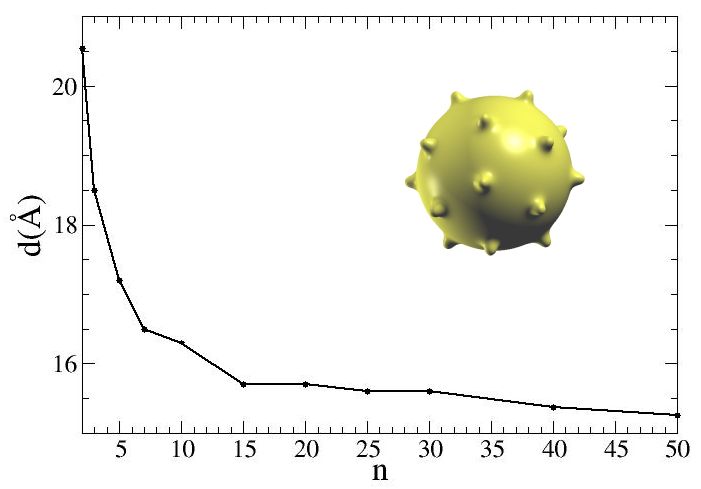}}
\caption{ 
Distance of the Na$^+$ ion to the droplet surface as a function of the number of charges $n$ for droplets of radius $R_{expl}$.
The line has been drawn to guide the eye. The inset cartoon represents qualitatively
the actual shape of a droplet with $n=20$ and a radius close to $R=R_{expl}$.
The density is approximately obtained by superimposing 
a finite-width He density profile mimicking the one in Fig. \ref{fig1} with gaussian profiles 
centered around each cation.  
}
\label{fig10}
\end{figure}

\section{Summary and concluding remarks}

We have designed a solvation plus Coulomb model for studying the stability of extrinsic multiply  charged droplets, 
and applied it within the He-DFT approach.
 We believe that our approach is  
an improvement over those based on the liquid drop model plus Rayleigh's stability criterion, 
 since it is  built on a more accurate description of the solvation 
properties of positive ions in helium.
In particular, it explicitly takes into account the ion-snowball microclusters 
which are eventually ejected after Coulomb explosion 
as found in the experiments.

 In addition to the stability study, we  have disclosed the existence of energy barriers 
  in the dissociation process of the ion microclusters from the helium droplet.
These barriers explain why multiple  ions-droplet systems can be observed}  for droplet sizes smaller than predicted for their
 stability: they are metastable, at least during the millisecond time range of the experiments. 

We have defined the critical radius for the  (meta)stability of multicharged droplets as that 
for which the energy barrier equals the thermal activation energy. 
This critical radius can be identified in practice with the radius $R_{expl}$
for which the charged droplet would undergo Coulomb explosion.

 We have found that close to the critical radius, charges are localized just above 
the helium surface, resulting in a virus-like appearance of the 
critical multicharged helium droplet.
For radii increasingly larger than the critical one, 
charges submerge but still remain close to the droplet surface.

It is interesting to note that the energy barriers result from the interplay 
between ion solvation and Coulomb energies.
This is at variance with nuclear fission barriers, where the interplay is between surface and Coulomb energies.\cite{Boh98}
Yet, a proper description of the droplet surface, such as the one provided by modern density functionals,\cite{Dal95,Anc05} 
was essential here to take surface tension correctly into account.
 
Considering  helium droplets doped with Na$^+$ or Ca$^+$ ions as study cases, 
we conclude that the results are rather species independent.
Hence  it is very tempting to compare our results with those
obtained for pristine drops,\cite{Lai19} even though the mechanisms for producing extrinsic and intrinsic multicharged helium droplets
 are  different.
 The results obtained here reproduce the linear dependence of the number of hosted ions with the square of the critical radius.
 However, the agreement is qualitative, our values for $R_{expl}$ being systematically smaller than the experimental critical radius
 for intrinsic multicharged helium droplets.\cite{Lai19}
The agreement is still better than the one found with the LDM plus Rayleigh stability criterion of a 
charged liquid droplet,\cite{Liv07}
which in addition does not reproduce the linear behaviour of $n$ vs $R^2$.
  
It is worth stressing that we have carried out static calculations 
to determine the energy barriers and critical radii.
In the real world, the process is dynamic: for
 intrinsic multiply charged helium droplets, He$_3^+$ ions are produced 
by electron impact  near the surface facing the electron source\cite{Lai19,Ell07}
and readjust under the Coulomb repulsion to locations similar to Thomson's 
model configurations in order to minimize the total energy. Some of these charges may have
enough kinetic energy to leave the droplet.
Consequently, 
the critical radius is expected to be larger than in our predictions based on static models   where
 charges are  supposed to be evenly distributed on the droplet surface.
However, considering an Atkins' snowball of 200 amu  mass\cite{Mat14}
moving at a maximum velocity of the order of Landau's velocity, say 50 m/s, its kinetic energy
is only about 30 K, so one should not expect a large effect on the critical radius. 
For instance, in the case of $n=2$ a barrier of 30 K appears in the 
$R=27.1$ \AA{} droplet, only 2.5 \AA{} larger than $R_{expl}$.

Even within our static approach, we can get a flavor of what the effect of the ionization dynamics could be.
Taking advantage of the fact that steepest-descent minimization usually yields the closest local minimum, 
we have obtained metastable configurations with an energy higher than that of the absolute minimum (the solution of Thomson's model)
by starting the SD procedure from a distribution of ions inside the bulk of a droplet.
For instance, if we start with $n=20$ ions distributed within 
a droplet of radius $R=350$ \AA{}  (large 
enough to guarantee the stability of the droplet-ions system), 
we end up with a surface distribution of
the positive ions but with
an energy higher by $\sim 730$ K than that of the solution of Thomson' model, which is $5.879\times 10^4$ K.
This metastable configuration is buckled, i.e., all the ions do not reside at the same distance from the droplet center:
the maximum difference is $\sim 4$ \AA.
The calculated value for the critical radius is then larger than the one obtained from the ideal Thomson configuration,
 $R_{expl}=254$~\AA{} instead of $221$~\AA{}, but it is still smaller than the experimental $R_c$ value of $ \sim 281$~\AA.\cite{Lai19}
 Interestingly, for a droplet with a radius slightly smaller than $254$~\AA, Coulomb explosion occurred differently from the symmetric dynamics of the ``ideal'' structure where all cations move  radially away from the droplet.
 For this metastable configuration, only one cation was expelled, and the remaining $19$ ones readjusted their positions and remained on the surface of the droplet.

Another possible source of  discrepancies between our 
calculations and experiments is our substitution 
of the droplet geometry by a planar surface  in order to compute the
solvation energy.
This would  mainly affect the predicted $R_{expl}$ values for small $n$ charged drops, where curvature effects are more important.

We acknowledge that it might not be easy 
to carry out an experiment on extrinsic multiply charged helium droplets 
obtained from droplets doped with alkali or the heavier alkaline earth 
atoms, due to their tendency to form clusters  and sink inside the droplet. This might
happen before the dopant atoms are ionized, hampering the formation of 
the initial multi-ion configuration object of this study.
Yet, we hope that our results on extrinsic multiply charge drops 
will motivate further experiments for determining if there is a quantitative difference
between them and those on intrinsic multiply charged drops and if yes, what the origin could be.

\appendix
\setcounter{equation}{0}
\section{Interaction pair potentials}

\begin{table}
\begin{center}
\begin{tabular}{|c|c|c|}

\hline
  & Ca$^+$-He &  He$_3^+$ -He \\
\hline
$g_1$       &  9.71136 $\times 10^6$   &  1.17674   $\times 10^7$   \\
$g_2$       &  -1.68041 $\times 10^7$ &  -1.77677  $\times 10^7$\\
$g_3$       & 1.12045  $\times 10^7$ & 1.15284 $\times 10^7$ \\                 
$g_4$       & -3.20735  $\times 10^6$ & -3.65381  $\times 10^6$\\
$g_5$       &  8.79979 $\times 10^4$ & 2.66532  $\times 10^5$ \\
$g_6$       & 1.71418  $\times 10^5$ &  1.41737 $\times 10^5$\\
$g_7$       & -3.10192  $\times 10^4$&  -2.19406 $\times 10^4$\\
$g_8$       &  -46.583  &  -555.796 \\
$g_9$       & 261.538  &  -138.371  \\
$\alpha$     & 3.443  &  6.109 \\
$\beta$     & -2.858  & -8.837  \\
$C_4$     & 1.54022  $\times 10^4$& 1.81590  $\times 10^4$ \\
$C_6$   &  3.37909   $\times 10^5$& 1.74591 $\times 10^4$ \\
$C_8$   & -1.73043 $\times 10^6$&   4.57316  $\times 10^5$ \\
\hline
\end{tabular}
\end{center}
\caption{ Parameters of the Ca$^+$-He and He$_3^+$-He pair potentials,
Eqs. (\ref{eqA2}) and (\ref{eqA3}).
\label{Table1}
}
\end{table}

The potential energy surface (PES) of the Ca$^+$-He system have been 
determined at the CCSD(T) level of theory, adopting high quality 
basis sets and a fine grid of internuclear distances. 
When performing first principles computations, we explicitly considered all 21 
electrons (19 belonging to Ca$^+$ and 2 to He) to evaluate the electron 
correlation contribution to the interaction energy. 
All data have been corrected for basis set superposition error 
using the well-known counterpoise procedure proposed by Boys and Bernardi.\cite{Boy70} 
As for the basis set, we adopted the all-electrons \mbox{Def2-QZVPPD}\cite{Rap10} 
for calcium, and the \mbox{aug-cc-pVQZ}\cite{Woo94} for helium. 
To approach the limit of basis set completeness, they have been 
supplemented with a \textit{3s3p2d} basis of bond functions\cite{Tao92} 
placed at midway between the Ca$^+$ and the He nuclei. 
To test the reliability of the basis set choice, we performed two  limited series of test computations. 
The electrons of helium have been described with the \mbox{aug-cc-pV5Z}\cite{Woo94} set 
as well as with the \mbox{aug-cc-pV6Z}\cite{Mou99} one. 
In both cases the relevant features of the PES undergo negligible variations. 
The Ca$^+$-He separation of the minimum interaction energy varies by
less than 0.01 \AA{}, and the same holds true for the turning point, 
i.e. the distance where the PES reverts from attractive to repulsive. 
Accordingly, the depth of the attractive well increases by 
about 0.5{\%}. Overall, data obtained with the three different basis sets for He are 
almost identical, whatever the internuclear distance considered. 
A test on the theoretical method has been conducted as well. 
Following a procedure well grounded in the case of weakly bound 
complexes,\cite{Hin03,Mar94} we evaluated the electron correlation 
contribution of the three outermost electrons (4s$^1$ belonging to Ca$^+$, 1s$^2$  
to He) at the Full-Configuration-Interaction (FCI) 
level of theory. To get a reliable interaction potential, the 
FCI results have been summed up to the contribution of the 18 inner electrons,
which has been evaluated with the CCSD(T) approach. 
Also in this case no relevant change in the PES has been recorded. 
We are therefore confident that our computational scheme is 
capable of describing the Ca$^+$-He complex with a high degree of accuracy. To
determine the entire PES, we defined a spatial grid consisting 
in 100 internuclear separations, such as to sample finely 
(i.e. with 0.10 \AA{} steps) the repulsive wall and the attractive well, 
while enlarging the grid in the long-range tail.

The overall properties of the PES can be summarized as follows: 
\textit{(i)} the depth of the attractive well amounts to 35.92 cm$^{-1}$; \textit{(ii)} 
the minimum interaction energy is located at the Ca$^+$-He distance of 4.27 \AA{}; 
\textit{(iii)} the PES reverts from being attractive to 
repulsive at 3.64 \AA{} 
and \textit{(iv)} the repulsive wall reaches {$10^3$} cm$^{-1}$ at 2.43 \AA{}. 
These data agree very well with a recent study on 
this same complex,\cite{Bar21} and they are consistent with available 
results obtained at a slightly lower level of theory.\cite{Czu96,Fie12}

The {\it ab initio} Ca$^+$-He pair potential has been fit to an analytical expression 
similar to that of Buchachenko et al.\cite{Bub05}
\begin{equation}
V(r) = V_{SR}(r) + V_{LR}(r) \, ,
\label{eqA1}
\end{equation}
where
\begin{equation}
V_{SR}(r) = \left[\sum_{i=1}^{9} \,g_i \,r^{(i-1)}\right] \, e^{(-\alpha r -\beta)}
\label{eqA2}
\end{equation}
represents a short-range repulsive interaction, and $V_{LR}$ corresponds to a long-range attractive interaction.
For ions,
\begin{equation}
V_{LR}(r) =  \frac{C_4}{r^4}+\frac{C_6}{r^6}+\frac{C_8}{r^8} 
\label{eqA3}
\end{equation}
The parameters were determined using a nonlinear least-squares algorithm. 
The value of the parameters is given in Table \ref{Table1}.
For the sake of completeness, we also give the  
parameters of the spherically averaged He$_3^+$-He pair potential
of Ref. \onlinecite{Mat14}, the raw data of which have been 
obtained by digitizing the curve in Fig. 4 of that reference.
Units of the parameters are such that when $r$ is given in \AA, $V(r)$ is obtained in K.
The {\it ab initio} points and fit are shown in Fig. \ref{fig4}.
The overall quality of the fits is very good.

 \begin{acknowledgments}
We thank Paul Scheier and Marcel Mudrich for useful discussions. 
A computer grant from  CALMIP high performance computer center (grant P1039) is gratefully acknowledged.
This work has been  performed under Grant No.  PID2020-114626GB-I00 from the MICIN/AEI/10.13039/501100011033
and benefited from COST Action CA21101 ``Confined molecular systems: 
from a new generation of materials to the stars'' (COSY) 
supported by COST (European Cooperation in Science and Technology).
\end{acknowledgments}

\bigskip

\section*{AUTHOR DECLARATIONS}   
\subsection*{Conflict of Interest}
The authors have no conflicts to disclose.

\subsection*{DATA AVAILABILITY}
The data that support the findings of this study are available
from the corresponding author upon reasonable request.

\end{document}